\documentclass{llncs}
\usepackage{amsmath,amssymb,amsfonts}
\usepackage{mathtools}
\usepackage[usenames,dvipsnames]{color}
\usepackage{xcolor}
\usepackage{xspace}
\usepackage{microtype}
\usepackage{todonotes}
\usepackage{colonequals}
 \usepackage{booktabs}

\usepackage{multirow}
\usepackage{multicol}

\usepackage{paralist}

\usepackage{subfigure}
\usepackage{url}
\usepackage{appendix}
\usepackage{graphicx}
\usepackage{algorithm}
\usepackage{listings}
\usepackage{nicefrac}
\usepackage{tikz} 
\usepackage{pgfplots}

\usetikzlibrary{arrows,decorations.pathmorphing,positioning,fit,trees,shapes,shadows,automata,calc} 

\tikzset{outline/.style args={#1}{%
  draw=#1,thick,fill=#1!50}}

\tikzset{
  dot hidden/.style={},
  line hidden/.style={},
  dice hidden/.style={},
  dot color/.style={dot hidden/.append style={color=#1}},
  dot color/.default=black,
  line color/.style={line hidden/.append style={color=#1}},
  line color/.default=black,
  dice color/.style={dice hidden/.append style={color=#1,fill}},
  dice color/.default=white
}\def\dotsize{0.1}
\newcommand{\drawdie}[2][]{%
\begin{tikzpicture}[x=1em,y=1em,#1]
  \draw 	[thick, rounded corners=0.5,line hidden,dice hidden] (0,0) rectangle (1,1);
  \ifodd#2
    \fill[dot hidden] (0.5,0.5) circle (\dotsize);
  \fi
  \ifnum#2>1
  \fill[dot hidden] (0.25,0.25) circle (\dotsize);
  \fill[dot hidden] (0.75,0.75) circle (\dotsize);
  \ifnum#2>3
    \fill[dot hidden] (0.25,0.75) circle (\dotsize);
    \fill[dot hidden] (0.75,0.25) circle (\dotsize);
    \ifnum#2>5
      \fill[dot hidden] (0.75,0.5) circle (\dotsize);
      \fill[dot hidden] (0.25,0.5) circle (\dotsize);
    \fi
  \fi
\fi
\end{tikzpicture}
}

\title{Sequential Convex Programming for the Efficient Verification of Parametric MDPs%
\thanks{Partly funded by the awards AFRL \# FA8650-15-C-2546, DARPA \# W911NF-16-1-0001,  
ARO \# W911NF-15-1-0592, ONR \#  N00014-15-IP-00052, ONR \# N00014-16-1-3165, and NSF \# 1550212. Also funded by the Excellence Initiative of the German
federal and state government and the CDZ project CAP (GZ 1023).}
}
\author{Murat Cubuktepe\inst{1}, Nils Jansen\inst{1}, Sebastian Junges\inst{2}, Joost-Pieter Katoen\inst{2},\\ Ivan Papusha\inst{1}, Hasan A. Poonawala\inst{1}, Ufuk Topcu\inst{1}}
\institute{The University of Texas at Austin, USA  \and RWTH Aachen University, Germany}

\newcommand{\ie}{i.e.\@\xspace}

\newcommand{\eg}{e.g.\@\xspace}

\newcommand{\prophesy}{\textrm{PROPhESY}\xspace}

\newcommand{\tool}[1]{\textrm{#1}\xspace}

\newcommand{\dtmc}{\mathcal{D}}

\newcommand{\p}{\ensuremath{\mathbb{P}}}
\newcommand{\pr}{\ensuremath{\mathrm{Pr}}}
\newcommand{\reachPr}[2]{\ensuremath{\pr^{#1}(\finally #2)}}
\newcommand{\reachPrT}[1][]{\ensuremath{\reachPr{#1}{T}}}
\newcommand{\reachProp}[2]{\ensuremath{\p_{\leq #1}(\finally #2)}}
\newcommand{\reachProplT}{\ensuremath{\reachProp{\lambda}{T}}}
\newcommand{\reachPrs}[3]{\ensuremath{\pr^{#1}_{#2}(\finally #3)}}
\newcommand{\reachPropSymbol}{\varphi}

\newcommand{\er}{\ensuremath{\mathrm{EC}}}

\newcommand{\expRewProp}[2]{\ensuremath{\er_{\leq #1}(\finally #2)}}

\newcommand{\rewFunction}{\ensuremath{{c}}}

\newcommand{\finally}{\lozenge}

\newcommand{\sj}[1]{}

\newcommand{\R}{\mathbb{R}}

\newcommand{\N}{\mathbb{N}}

\newcommand{\Ireal}{[0,\, 1]\subseteq\mathbb{R}}  

\newcommand{\Distr}{\mathit{Distr}}

\newcommand{\distDom}{X}

\newcommand{\distFunc}{\mu}
\newcommand{\distDomElem}{x}

\newcommand{\Var}{\ensuremath{\mathcal{V}}\xspace}        
\newcommand{\Paramvar}{\ensuremath{{V}}\xspace}        

\newcommand{\State}{\ensuremath{\mathbb{S}}\xspace}      

\newcommand{\sinit}{s_{\mathit{I}}} 
\newcommand{\mdp}{\mathcal{M}}
\newcommand{\MdpInit}[1][]{\ensuremath{\mdp{#1}=(S{#1},\sinit{#1},\Act,\probmdp{#1})}}
\newcommand{\pMdpInit}[1][]{\ensuremath{\mdp{#1}=(S{#1},\,\sinit{#1},\Act,\Paramvar,\probmdp{#1})}}
\newcommand{\probmdp}{\mathcal{P}}

\newcommand{\posy}{\ensuremath{f}}
\newcommand{\mono}{\ensuremath{g}}

\newcommand{\monos}[1][]{\ensuremath{\mathit{Mon_{#1}}}}
\newcommand{\posys}[1][]{\ensuremath{\mathit{Pos_{#1}}}}
\newcommand{\signomis}[1][]{\ensuremath{\mathit{Sig_{#1}}}}

\newcommand{\valuations}{\ensuremath{\mathit{Val}}}

\newcommand{\sched}{\ensuremath{\sigma}}
\newcommand{\Sched}{\ensuremath{\mathit{Sched}}}

\newcommand{\Act}{\ensuremath{\mathit{Act}}}
\newcommand{\act}{\ensuremath{\alpha}}
\newcommand{\pmdp}{\ensuremath{\mathcal{P}}}

\DeclareMathAlphabet{\mathpzc}{OT1}{pzc}{m}{it}
\def\presuper#1#2%
  {\mathop{}%
   \mathopen{\vphantom{#2}}^{#1}%
   \kern-\scriptspace%
   #2}

\begin{document}

\maketitle
\begin{abstract}
Multi-objective verification problems of parametric Markov decision
processes under optimality criteria can be naturally expressed as
nonlinear programs. We observe that many of these computationally
demanding problems belong to the subclass of signomial programs.
This insight allows for a sequential optimization algorithm to
efficiently compute sound but possibly suboptimal solutions. Each stage
of this algorithm solves a geometric programming problem. These geometric
programs are obtained by convexifying the nonconvex constraints of the
original problem. 
Direct applications of the encodings as nonlinear programs are model repair and parameter synthesis.
We demonstrate the scalability and quality of our
approach by well-known benchmarks.
\end{abstract}

\section{Introduction}
\label{sec:introduction}
We study the applicability of \emph{convex optimization} to the formal verification of systems that exhibit randomness or stochastic uncertainties. Such systems are formally represented by so-called parametric Markov models. 

In fact, many real-world systems exhibit random behavior and stochastic uncertainties. One major example is in the field of \emph{robotics}, where the presence of measurement noise or input disturbances requires special controller synthesis techniques~\cite{thrun2005probabilistic} that achieve robustness of robot actions against uncertainties in the robot model and the environment.
On the other hand, formal verification offers methods for rigorously proving or disproving properties about the system behavior, and synthesizing strategies that satisfy these properties.
In particular, \emph{model checking}~\cite{BK08} is a well-studied technique that provides guarantees on appropriate behavior for all possible events and scenarios.

Model checking can be applied to systems with stochastic uncertainties, including discrete-time Markov chains (MCs), Markov decision processes (MDPs), and their continuous-time counterparts~\cite{katoen2016probabilistic}.
Probabilistic model checkers are able to verify reachability properties like ``the probability of reaching a set of unsafe states is $\leq 10\%$'' and expected costs properties like ``the expected cost of reaching a goal state is $\leq 20$.''
A rich set of properties, specified by linear- and branching-time logics, reduces to such properties~\cite{katoen2016probabilistic}.
Tools like \tool{PRISM}~\cite{KNP11}, \tool{STORM}~\cite{DBLP:journals/corr/DehnertJK016}, and 
\tool{iscasMc}~\cite{iscasmc} are probabilistic model checkers capable of handling a wide range of large-scale problems.

Key requirements for applying model checking are a reliable system model and formal specifications of desired or undesired behaviors. 
As a result, most approaches assume that models of the stochastic uncertainties are precisely given. For example, if a system description includes an environmental disturbance, the mean of that disturbance should be known \emph{before} formal  statements are made about expected system behavior.
However, the desire to treat many applications where uncertainty measures (\eg, faultiness, reliability, reaction rates, packet loss ratio) are not exactly known at design time 
gives rise to \emph{parametric} probabilistic models~\cite{DBLP:journals/ior/SatiaL73,DBLP:journals/ai/DelgadoBDS16}. Here, transition probabilities are expressed as functions over system parameters, \ie, \emph{descriptions of uncertainties}.
In this setting, \emph{parameter synthesis} addresses the problem of computing parameter instantiations leading to satisfaction of system specifications. 
More precisely, parameters are mapped to concrete probabilities inducing the resulting \emph{instantiated} model to satisfy  specifications.
A direct application is \emph{model repair}~\cite{bartocci2011model}, where a concrete model (without parameters) is changed (repaired) such that specifications \emph{are} satisfied.
%
%

Dedicated tools like \tool{PARAM}~\cite{PARAM10}, \tool{PRISM}~\cite{KNP11}, or \prophesy~\cite{dehnert-et-al-cav-2015} compute rational functions over parameters that express reachability probabilities or expected costs in a parametric Markov chain (pMC). These optimized tools work with millions of states but are restricted to a few parameters, as the necessary computation of greatest common divisors does not scale well with the number of parameters.
	Moreover, the resulting functions are inherently \emph{nonlinear} and often of high degree. Evaluation by an SMT solver over nonlinear arithmetic such as \tool{Z3}~\cite{demoura_nlsat} suffers from the fact that the solving procedures are \emph{exponential in the degree of polynomials and the number of variables}. 

This paper takes an alternative perspective. 
We discuss a general nonlinear programming formulation for the verification of parametric Markov decision processes (pMDPs).
 The powerful modeling capabilities of nonlinear programs (NLPs) enable incorporating multi-objective properties and penalties on the parameters of the pMDP.
However, because of their generality, solving NLPs to find a global optimum is difficult. Even feasible solutions (satisfying the constraints) cannot always be computed efficiently~\cite{bertsekas1999nonlinear,Las01}. 
In contrast, for the class of NLPs called \emph{convex optimization} problems, efficient methods to compute feasible solutions and global optima even for large-scale problems are available~\cite{boyd_convex_optimization}. 

We therefore propose a novel automated method of utilizing convex optimization for pMDPs.
%
Many NLP problems for pMDPs belong to the class of \emph{signomial programs} (SGPs), a certain class of nonconvex optimization problems.
For instance, all benchmarks available at the \tool{PARAM}--webpage~\cite{param_website} belong to this class.
%
%
Restricting the general pMDP problem accordingly yields a direct and efficient synthesis method---formulated as an NLP---for a large class of pMDP problems. 
 We list the two main technical results of this paper:
\begin{enumerate}
	\item We relax nonconvex constraints in SGPs and apply a simple transformation to the parameter functions. The resulting programs are \emph{geometric programs} (GPs)~\cite{boyd2007tutorial}, a class of \emph{convex programs}. We show that a solution to the relaxed GP induces feasibility (satisfaction of all specifications) in the original pMDP problem. Note that solving GPs is \emph{polynomial} in the number of variables. 
	\smallskip
	\item Given an initial feasible solution, we use a technique called \emph{sequential convex programming}~\cite{boyd2007tutorial} to improve a signomial objective. This local optimization method for nonconvex problems leverages convex optimization by solving a sequence of convex approximations (GPs) of the original SGP. \end{enumerate}
Sequential convex programming is known to efficiently find a feasible solution with
good, though not necessarily globally optimal, objective values~\cite{boyd2007tutorial,boyd2008sequential}.
We initialize the sequence with a feasible solution (obtained from the GP) of the original problem and compute a \emph{trust region}. Inside this region, the optimal value of the approximation of the SGP is at least as good as the objective value at the feasible solution of the GP. 
The optimal solution of the approximation is then the initial point of the next iteration with a new trust region.
This procedure is iterated to approximate a local optimum of the original problem. 
%

Utilizing our results, we discuss the concrete problems of parameter synthesis and model repair for multiple specifications for pMDPs.
Experimental results with a prototype implementation show the applicability of our optimization methods to benchmarks of up to $10^5$ states.
As solving GPs is polynomial in the number of variables, our approaches are relatively insensitive to the number of parameters in pMDPs. This is an improvement over state-of-the-art approaches that leverage SMT, which---for our class of problems---scale exponentially in variables and the degree of polynomials. 
This is substantiated by our experiments.

\paragraph{Related work.}
Several approaches exist for pMCs~\cite{PARAM10,dehnert-et-al-cav-2015,param_sttt,jansen-et-al-qest-2014} while the number of approaches for pMDPs~\cite{param_sttt,quatmann-et-al-atva-2016} is limited.
Ceska \emph{et al.}~\cite{DBLP:conf/cmsb/CeskaDKP14} synthesize rate parameters in stochastic biochemical networks. 
Multi-objective model checking of non-parametric MDPs~\cite{DBLP:journals/lmcs/EtessamiKVY08} is a convex problem~\cite{DBLP:conf/tacas/ForejtKNPQ11}. 
Bortolussi \emph{et al.}~\cite{DBLP:journals/iandc/BortolussiMS16} developed a Bayesian statistical algorithm for properties on stochastic population models.
Convex uncertainties in MDPs without parameter dependencies are discussed in~\cite{seshia_et_al_cav_13}.
Parametric probabilistic models are used to rank patches in the repair of software~\cite{DBLP:conf/popl/LongR16} and to compute perturbation bounds~\cite{rosenblum-et-al-atva-2014,su-et-al-icse-2016-qosevaluation}.

\section{Preliminaries}
\label{sec:preliminaries}
%


%
A \emph{probability distribution} over a finite or countably infinite set $\distDom$ is a function $\distFunc\colon\distDom\rightarrow\Ireal$ with $\sum_{\distDomElem\in\distDom}\distFunc(\distDomElem)=1$. 
The set of all distributions on $\distDom$ is denoted by $\Distr(\distDom)$.

%
\begin{definition}[Monomial, Posynomial, Signomial]\label{def:posy}
  Let $V=\{x_1,\ldots,x_n\}$ be a finite set of strictly positive real-valued \emph{variables}.
  A \emph{monomial} over $V$ is an expression of the form
  \begin{align*}
      \mono=c\cdot x_{1}^{a_{1}}\cdots x_{n}^{a_{n}}\ ,
  \end{align*}
  where $c\in \R_{>0} $ is a positive coefficient, and $a_i\in\R$ are exponents for $1\leq i\leq n$. 
  A \emph{posynomial} over $V$ is a sum of one or more monomials:
  \begin{align}
      \posy=\sum_{k=1}^K c_k\cdot x_1^{a_{1k}}\cdots x_n^{a_{nk}} \ .\label{eq:signomial}
  \end{align}
  If $c_k$ is allowed to be a negative real number for any $1\leq k\leq K$, then the expression~\eqref{eq:signomial} is a \emph{signomial}. The sets of all monomials, posynomials, and signomials over $V$ are denoted by $\monos[V]$, $\posys[V]$, and $\signomis[V]$, respectively.
 \end{definition}
This definition of monomials differs from the standard algebraic definition where exponents are positive integers with no restriction on the coefficient sign. A sum of monomials is then called a \emph{polynomial}. Our definitions are consistent with~\cite{boyd2007tutorial}.
\begin{definition}[Valuation]
  \label{def:valuation}
  For a set of real-valued variables $V$, a \emph{valuation $u$ over $V$} is a function $u\colon V \rightarrow \R$.
  The set of all valuations over $V$ is $\valuations^V$.
\end{definition}
 Applying valuation $u$ to monomial $\mono$ over $V$ yields a real number $\mono[u]\in\R$ by replacing each occurrence of variables $x\in V$ in $\mono$ by $u(x)$; the procedure is analogous for posynomials and signomials using standard arithmetic operations.


 %
%
\begin{definition}[pMDP and pMC]\label{def:pmdp}
A \emph{parametric Markov decision process (pMDP)} is a tuple $\pMdpInit$ with a finite set $S$ of states, an initial state $\sinit \in S$, a finite set $\Act$ of actions, a finite set of real-valued variables $\Paramvar$, and a transition function $\probmdp \colon S \times \Act \times S \rightarrow \signomis[\Paramvar]$ satisfying for all $s\in S\colon
\Act(s) \neq \emptyset$,  where $\Act(s) = \{\act \in \Act \mid \exists s'\in S.\,\probmdp(s,\,\act,\,s') \neq 0\}$.
If for all $s\in S$ it holds that $|\Act(s)| = 1$, $\mdp$ is called a \emph{parametric discrete-time Markov chain (pMC)}
.
\end{definition}
%
%
%
$\Act(s)$ is the set of \emph{enabled} actions at state $s$; as $\Act(s) \neq \emptyset$, there are no deadlock states.
%
\emph{Costs} are defined using a state--action \emph{cost function} $\rewFunction \colon S \times \Act \rightarrow \R_{\geq 0}$.
\begin{remark}
Largely due to algorithmic reasons, the transition probabilities in the literature~\cite{param_sttt,dehnert-et-al-cav-2015,quatmann-et-al-atva-2016} are polynomials or rational functions, \ie, fractions of polynomials.  
 Our restriction to signomials is realistic; \emph{all} benchmarks from the \tool{PARAM}--webpage~\cite{param_website} contain only signomial transition probabilities. 
\end{remark}

A pMDP $\mdp$ is a \emph{Markov decision process (MDP)} if the transition function is a valid probability distribution, \ie, $\probmdp \colon S \times \Act \times S \rightarrow [0,1]$ and $\sum_{s'\in S}\probmdp(s,\act,s') = 1$ for all $s \in S \mbox{ s.t.\ } \act \in \Act(s)$. 
Analogously, a Markov chain (MC) is a special class of a pMC; a model is \emph{parameter-free} if all  probabilities are constant. Applying a \emph{valuation} $u$ to a pMDP, denoted $\mdp[u]$, replaces each signomial $f$ in $\mdp$ by $f[u]$; we call $\mdp[u]$ the \emph{instantiation} of $\mdp$ at $u$.
The application of $u$ is to replace the transition function $f$ by the probability $f[u]$. 
A valuation $u$ is \emph{well-defined} for $\mdp$ if the replacement yields \emph{probability distributions} at all states; the resulting model $\mdp[u]$ is an MDP or an MC.

\begin{figure}[t]
\centering
	\subfigure[pMC model]{
		\scalebox{0.75}{
			\begin{tikzpicture}[scale=1, die/.style={inner sep=0,outer sep=0},nodestyle/.style={draw,circle},baseline=(s0)]
    
    \node [nodestyle] (s0) at (0,0) {$s_0$};
    \node [] (leftdummy)  [on grid, left=1.2cm of s0] {};
    \node [] (rightdummy) [on grid, right=1.2cm of s0] {};
    \node [nodestyle] (s1) [on grid, below=1.1cm of leftdummy] {$s_1$};
    \node [nodestyle] (s2) [on grid, below=1.1cm of rightdummy] {$s_2$};
    \node [nodestyle] (s3) [on grid, below=1.3cm of s1, xshift=-0.5cm] {$s_3$};
        \node [nodestyle] (s4) [on grid, below=1.3cm of s1, xshift=0.5cm] {$s_4$};
    \node [nodestyle] (s5) [on grid, below=1.3cm of s2, xshift=-0.5cm] {$s_5$};
    
    \node [nodestyle] (s6) [on grid, below=1.3cm of s2, xshift=0.5cm] {$s_6$};

    \node[die, scale=1.5, below=0.6cm of s3, xshift=-0.3cm] (X1) {\drawdie{1}};
    \node[die, scale=1.5, right=0.21cm of X1, inner sep=0pt] (X2) {\drawdie{2}};
    \node[die, scale=1.5, right=0.21cm of X2] (X3) {\drawdie{3}};
    \node[die, scale=1.5, right=0.21cm of X3] (X4) {\drawdie{4}};
    \node[die, scale=1.5, right=0.21cm of X4] (X5) {\drawdie{5}};
    \node[die, scale=1.5, right=0.21cm of X5] (X6) {\drawdie{6}};
    
    \draw ($(s0)-(0.7,0)$) edge[->] (s0);
    \draw (s0) edge[->] node[right] {\scriptsize$p$} (s1);
    \draw (s0) edge[->] node[right] {\scriptsize$1{-}p$} (s2);
    \draw (s1) edge[bend left, ->] node[left] {\scriptsize$q$} (s3);
    \draw (s1) edge[->] node[right] {\scriptsize$1{-}q$} (s4);
    \draw (s3) edge[bend left, ->] node[left] {\scriptsize$p$} (s1);
    \draw (s3) edge[->] node[left] {\scriptsize$1{-}p$} (X1);
    \draw (s4) edge[->] node[left] {\scriptsize$1{-}p$} (X2);
    \draw (s4) edge[->] node[right] {\scriptsize$p$} (X3);
    \draw (s2) edge[bend left, ->] node[left] {\scriptsize$q$} (s5);
    \draw (s2) edge[->] node[right] {\scriptsize$1{-}q$} (s6);
    \draw (s5) edge[bend left, ->] node[left] {\scriptsize$p$} (s2);
    \draw (s5) edge[->] node[left] {\scriptsize$1{-}p$} (X4);
    \draw (s6) edge[->] node[left] {\scriptsize$1{-}p$} (X5);
    \draw (s6) edge[->] node[right] {\scriptsize$p$} (X6);
    
    \node[draw=white, rectangle, fit=(current bounding box), inner sep=3pt, outer sep=0pt] {};
    
\end{tikzpicture}
	\label{fig:pkydie}
		}
	}
	\subfigure[Instantiation using $p{=}0.4$ and $q{=}0.7$]{
		\scalebox{0.75}{
			\begin{tikzpicture}[scale=1, die/.style={inner sep=0,outer sep=0},nodestyle/.style={draw,circle},baseline=(s0)]
    
    \node [nodestyle] (s0) at (0,0) {$s_0$};
    \node [] (leftdummy)  [on grid, left=1.2cm of s0] {};
    \node [] (rightdummy) [on grid, right=1.2cm of s0] {};
    \node [nodestyle] (s1) [on grid, below=1.1cm of leftdummy] {$s_1$};
    \node [nodestyle] (s2) [on grid, below=1.1cm of rightdummy] {$s_2$};
    \node [nodestyle] (s3) [on grid, below=1.3cm of s1, xshift=-0.5cm] {$s_3$};
        \node [nodestyle] (s4) [on grid, below=1.3cm of s1, xshift=0.5cm] {$s_4$};
    \node [nodestyle] (s5) [on grid, below=1.3cm of s2, xshift=-0.5cm] {$s_5$};
    
    \node [nodestyle] (s6) [on grid, below=1.3cm of s2, xshift=0.5cm] {$s_6$};

    \node[die, scale=1.5, below=0.6cm of s3, xshift=-0.3cm] (X1) {\drawdie{1}};
    \node[die, scale=1.5, right=0.21cm of X1, inner sep=0pt] (X2) {\drawdie{2}};
    \node[die, scale=1.5, right=0.21cm of X2] (X3) {\drawdie{3}};
    \node[die, scale=1.5, right=0.21cm of X3] (X4) {\drawdie{4}};
    \node[die, scale=1.5, right=0.21cm of X4] (X5) {\drawdie{5}};
    \node[die, scale=1.5, right=0.21cm of X5] (X6) {\drawdie{6}};
    
    \draw ($(s0)-(0.7,0)$) edge[->] (s0);
    \draw (s0) edge[->] node[right] {\scriptsize$0.4$} (s1);
    \draw (s0) edge[->] node[right] {\scriptsize$0.6$} (s2);
    \draw (s1) edge[bend left, ->] node[left] {\scriptsize$0.7$} (s3);
    \draw (s1) edge[->] node[right] {\scriptsize$0.3$} (s4);
    \draw (s3) edge[bend left, ->] node[left] {\scriptsize$0.4$} (s1);
    \draw (s3) edge[->] node[left] {\scriptsize$0.6$} (X1);
    \draw (s4) edge[->] node[left] {\scriptsize$0.6$} (X2);
    \draw (s4) edge[->] node[right] {\scriptsize$0.4$} (X3);
    \draw (s2) edge[bend left, ->] node[left] {\scriptsize$0.7$} (s5);
    \draw (s2) edge[->] node[right] {\scriptsize$0.3$} (s6);
    \draw (s5) edge[bend left, ->] node[left] {\scriptsize$0.4$} (s2);
    \draw (s5) edge[->] node[left] {\scriptsize$0.6$} (X4);
    \draw (s6) edge[->] node[left] {\scriptsize$0.6$} (X5);
    \draw (s6) edge[->] node[right] {\scriptsize$0.4$} (X6);
    
    \node[draw=white, rectangle, fit=(current bounding box), inner sep=3pt, outer sep=0pt] {};
\end{tikzpicture}
	\label{fig:pkydiei}
		}
	}
\caption{A variant of the Knuth--Yao die for unfair coins.}
\vspace{-0.5cm}
\end{figure}
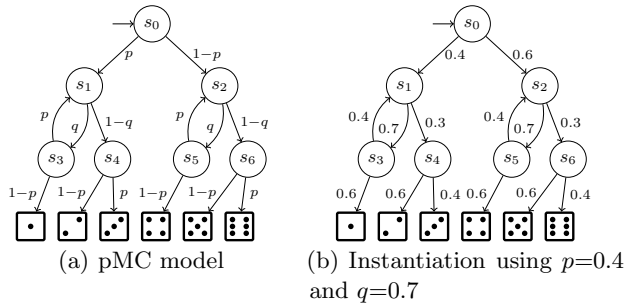

\begin{example}[pMC]\label{ex:die}
Consider a variant of the Knuth--Yao model of a die~\cite{KY76}, where a six-sided die is simulated by successive coin flips. 
We alternate flipping two biased coins, which result in \emph{heads} with probabilities defined by the monomials $p$ and $q$, respectively. Consequently, the probability for \emph{tails} is given by the signomials $1-p$ and $1-q$, respectively. The corresponding pMC is depicted in Fig.~\ref{fig:pkydie}; and the \emph{instantiated} MC for $p = 0.4$ and $q = 0.7$ is given in Fig.~\ref{fig:pkydiei}. Note that we omit actions, as the model is deterministic.
\end{example}
%
%
In order to define a probability measure and expected cost on MDPs, nondeterministic choices are resolved by so-called \emph{schedulers}.
For practical reasons we restrict ourselves to \emph{memoryless} schedulers; details can be found in~\cite{BK08}.
%

\begin{definition}[Scheduler]\label{def:scheduler}
	A (randomized) \emph{scheduler} for an MDP $\mdp$ is a function $\sched\colon S\rightarrow\Distr(\Act)$ such that $\sigma(s)(\alpha) > 0$ implies $\alpha \in \Act(s)$.
	The set of all schedulers over $\mdp$ is denoted by $\Sched^\mdp$.
\end{definition}
%
Applying a scheduler to an MDP yields a so-called \emph{induced Markov chain}.

\begin{definition}[Induced MC]\label{def:induced_dtmc}
	Let MDP $\MdpInit$ and scheduler $\sched\in\Sched^\mdp$. The \emph{MC induced by $\mdp$ and $\sched$} is  $\mdp^\sched=(S,\sinit,\Act,\pmdp^\sched)$ where for all $s,s'\in S$,
	\begin{align*}
		 \pmdp^\sched(s,s')=\sum_{\alpha\in\Act(s)} \sched(s)(\act)\cdot\pmdp(s,\alpha,s').
	\end{align*} 
\end{definition}
%
We consider \emph{reachability properties} and \emph{expected cost properties}.
For MC $\dtmc$ with states $S$, let $\reachPrs{\dtmc}{s}{T}$ denote the probability of reaching a set of \emph{target states} $T \subseteq S$ from  state $s\in S$; simply $\reachPrT[\dtmc]$ denotes the probability for initial state $\sinit$.
We use the standard probability measure as in~\cite[Ch.\ 10]{BK08}.
For threshold $\lambda\in [0,1]$, the \emph{reachability property} asserting that a target state is to be reached with probability at most $\lambda$ is denoted $\reachPropSymbol = \reachProplT$.
The property is satisfied by $\dtmc$, written $\dtmc \models \reachPropSymbol$, iff $\reachPrT[\dtmc]\leq\lambda$.

The cost of a path through MC $\dtmc$ until a set of \emph{goal states} $G\subseteq S$ is the sum of action costs visited along the path. The expected cost of a finite path is the product of its probability and its cost.
For $\reachPr{\dtmc}{G}  = 1$, the expected cost of reaching $G$ is the sum of expected costs of all paths leading to $G$.
An expected cost property $\expRewProp{\kappa}{G}$ is satisfied if the expected cost of reaching $T$ is bounded by a threshold $\kappa \in \R$.
Formal definitions are given in e.g.,~\cite{BK08}.

If multiple specifications $\varphi_1,\ldots,\varphi_q$ are given, which are either reachability properties or expected cost properties of the aforementioned forms, we write the satisfaction of all specifications $\varphi_1,\ldots,\varphi_q$ for an MC $\dtmc$ as $\dtmc\models\varphi_1\land\ldots\land\varphi_q$. 

An MDP $\mdp$ satisfies the specifications $\varphi_1,\ldots,\varphi_q$, iff \emph{for all} schedulers $\sched\in\Sched^\mdp$ it holds that $\mdp^\sched\models\varphi_1\wedge\ldots\wedge\varphi_q$. The verification of multiple specifications is also referred to as \emph{multi-objective model checking}~\cite{DBLP:journals/lmcs/EtessamiKVY08,DBLP:conf/atva/ForejtKP12}.
We are also interested in the so-called scheduler \emph{synthesis problem}, where the aim is to find a scheduler $\sched$ such that the specifications are satisfied (although other schedulers may not satisfy the specifications).

\section{Nonlinear programming for pMDPs}\label{sec:nonlinear}
In this section we formally state a general pMDP parameter synthesis problem and describe how
it can be formulated using nonlinear programming.

\subsection{Formal problem statement}\label{sec:problem} 
\fbox{
\begin{minipage}{\dimexpr\textwidth-4\fboxsep}
\begin{problem}\label{prob:pmdpsyn}
Given a pMDP $\pMdpInit$, specifications
$\varphi_1,\ldots,\varphi_q$ that are either probabilistic reachability
properties or expected cost properties, and an objective function
$f\colon\Paramvar\rightarrow\R$ over the variables $V$, compute a well-defined
valuation $u\in\valuations^V$ for $\mdp$, and a (randomized) scheduler $\sched\in\Sched^\mdp$
such that the following conditions hold:
\begin{enumerate}[(a)]
	\item \label{prob:pmdpsyn_a} \emph{Feasibility}:
		the Markov chain $\mdp^\sched[u]$ induced by scheduler $\sched$ and
		instantiated by valuation $u$ satisfies the specifications, \ie, $\mdp^\sched[u]\models\varphi_1 \wedge \ldots \wedge \varphi_q$.
	
	\item \label{prob:pmdpsyn_b} \emph{Optimality}:
		the objective $f$ is minimized.
\end{enumerate}
\end{problem}
\end{minipage}
}

\noindent Intuitively, we wish to compute a parameter valuation and a scheduler such
that all specifications are satisfied, and the objective is globally minimized.
We refer to a valuation--scheduler pair $(u,\sched)$ that satisfies
condition~(\ref{prob:pmdpsyn_a}), \ie, only guarantees satisfaction of the
specifications but does not necessarily minimize the objective $f$, as a
\emph{feasible} solution to the pMDP synthesis problem. If both
(\ref{prob:pmdpsyn_a}) and (\ref{prob:pmdpsyn_b}) are satisfied, the pair
is an \emph{optimal} solution to the pMDP synthesis problem.

\subsection{Nonlinear encoding}
We now provide an NLP encoding of Problem~\ref{prob:pmdpsyn}. A general NLP over 
a set of real-valued variables $\Var$ can be written as
\begin{align}
	\text{minimize} 		&\quad f\label{eq:nl_obj} \\
	\text{subject to} 		&\notag\\
	\forall i.\, 1\leq i\leq m 	&\quad g_i \leq 0,\label{eq:nl_ineq}\\
	\forall j.\, 1\leq i\leq p 	&\quad h_j = 0,\label{eq:nl_eq}
\end{align}
where $f$, $g_i$, and $h_j$ are arbitrary functions over $\Var$, and $m$ and $p$ are the 
number of inequality and equality constraints of the program respectively. Tools like
\tool{IPOPT}~\cite{ipopt} solve small instances of such problems.

Consider a
pMDP $\pMdpInit$ with specifications $\varphi_1=\reachProplT$ and $\varphi_2=\expRewProp{\kappa}{G}$. We will discuss how additional specifications of either type can be encoded.
The set $\Var = \Paramvar \cup W$ of variables of the NLP consists of
the variables $\Paramvar$ that occur in the pMDP as well as a set $W$ of additional variables:
\begin{itemize}
	\item $\{ \sched^{s,\alpha} \mid s \in S, \act\in\Act(s) \}$,
		which define the randomized scheduler $\sched$ by $\sched(s)(\act)=\sched^{s,\act}$.
	\item $\{ p_s \mid s \in S \}$, 
		where $p_s$ is the probability of reaching the target set 
		$T\subseteq S$ from state $s$ under scheduler $\sched$, and
	\item $\{ c_s \mid s \in S \}$, where $c_s$ is the expected cost to reach $G\subseteq S$ from $s$ under $\sched$.
\end{itemize}
 A valuation over $\Var$ consists of a valuation $u\in\valuations^V$ over the
pMDP variables and a valuation $w\in\valuations^W$ over the additional variables.
\begin{align}
	\mbox{minimize } &\quad f \label{eq:min_rand}\\
	\mbox{subject to}\notag \\
					 &\quad p_{\sinit}\leq \lambda,				\label{eq:lambda}\\
					 &\quad c_{\sinit}\leq \kappa,				\label{eq:kappa}\\
	\forall s\in S.	&\quad \sum_{\act\in\Act(s)}\sched^{s,\act}=1, \label{eq:well-defined_sched_rand}\\
	\forall s\in S\,~\forall\act\in\Act(s). &\quad 0 \leq \sched^{s,\act} \leq 1,				\label{eq:sched_is_dist}\\
	\forall s\in S\,~\forall\act\in\Act(s).	 &\quad \sum_{s'\in S}\probmdp(s,\act,s')=1,	\label{eq:well-defined_probs_rand}\\
	\forall s, s'\in S\,~\forall\act\in\Act(s).	 &\quad 0 \leq \probmdp(s,\act,s') \leq 1,	\label{eq:probs_is_prob}\\
	\forall s\in T.	&\quad p_s=1,															\label{eq:targetprob_rand}\\
	\forall s\in S\setminus T. &\quad p_s=\sum_{\act\in\Act(s)}\sigma^{s,\act}\cdot\sum_{s'\in S}	\probmdp(s,\act,s')\cdot p_{s'}, \label{eq:probcomputation_rand}\\
	\forall s\in G.	 &\quad c_s=0,															\label{eq:targetrew}\\
	\forall s\in S\setminus G.	&\quad c_s= \sum_{\act\in\Act(s)} \sigma^{s,\act} \cdot \Bigl(c(s,\act) + \sum_{s'\in S}	\probmdp (s,\act,s') \cdot c_{s'}\Bigr). \label{eq:rewcomputation}
\end{align}%
The NLP~\eqref{eq:min_rand}--\eqref{eq:rewcomputation} encodes Problem~\ref{prob:pmdpsyn} in the following way.
The objective function $f$ in~\eqref{eq:min_rand} is any real-valued function over the variables $\Var$. 
The constraints~\eqref{eq:lambda} and~\eqref{eq:kappa} encode the
specifications $\varphi_1$ and $\varphi_2$, respectively.
The constraints~\eqref{eq:well-defined_sched_rand}--\eqref{eq:sched_is_dist}
ensure that the scheduler obtained is well-defined by requiring that the
scheduler variables at each state sum to unity. 
Similarly, the constraints
\eqref{eq:well-defined_probs_rand}--\eqref{eq:probs_is_prob} ensure that
for all states, parameters from $\Paramvar$ are instantiated such that
probabilities sum up to one. 
(These constraints are included if not all probabilities at a state are constant.)
The probability of reaching the target for all states in the target set is
set to one using~\eqref{eq:targetprob_rand}.
The reachability probabilities in each state 
depend on the reachability of the successor states and the transition
probabilities to those states through~\eqref{eq:probcomputation_rand}.
Analogously to the reachability probabilities, the cost for each goal state $G\subseteq S$
must be zero, thereby precluding the collection of infinite cost at
absorbing states, as enforced by~\eqref{eq:targetrew}.
Finally, the expected cost for all states except target states is given by
the equation~\eqref{eq:rewcomputation}, where according to the
strategy $\sched$ the cost of each action is added to the expected cost of
the successors. 

	
We can readily extend the NLP to include more specifications. If
another reachability property $\varphi'=\reachProp{\lambda'}{T'}$ is given, we add the set of probability variables $\{ p'_s \mid
s \in S\}$ to $W$, and duplicate the 
constraints~\eqref{eq:targetprob_rand}--\eqref{eq:probcomputation_rand} accordingly.
To ensure satisfaction of $\varphi'$, we also add the constraint
$p'_{\sinit}\leq \lambda'$.
The procedure is similar for additional expected cost properties. 
By construction, we have the following result relating the NLP encoding and Problem~\ref{prob:pmdpsyn}.
\begin{theorem}
	\label{thm:soundcomplete}
	The NLP~\eqref{eq:min_rand}--\eqref{eq:rewcomputation} is sound and complete with respect to Problem~\ref{prob:pmdpsyn}.
\end{theorem}
We refer to soundness in the sense that each variable assignment that satisfies
the constraints induces a scheduler and a valuation of parameters such that a
feasible solution of the problem is induced. Moreover, any optimal solution to
the NLP induces an optimal solution of the problem. Completeness means that all
possible solutions of the problem can be encoded by this NLP; while
unsatisfiability means that no such solution exists, making the problem
\emph{infeasible}.

\paragraph{Signomial programs.} By Def.~\ref{def:posy} and~\ref{def:pmdp}, all constraints in the NLP consist of signomial functions.
A special class of NLPs known as \emph{signomial programs} (SGPs) is of the form~\eqref{eq:nl_obj}--\eqref{eq:nl_eq} where $f$, $g_i$ and $h_j$ are signomials over $\Var$, see Def.~\ref{def:posy}. Therefore, we observe that the NLP~\eqref{eq:min_rand}--\eqref{eq:rewcomputation} is an SGP. We will refer to the NLP as an SGP in what follows.

SGPs with equality constraints consisting of functions that are \emph{not affine} are not \emph{convex} in general. 
In particular, the SGP~\eqref{eq:min_rand}--\eqref{eq:rewcomputation} is not necessarily convex. Consider a simple pMC only having transition probabilities of the form $p$ and $1-p$, as in Example~\ref{ex:die}. The function in the equality
constraint~\eqref{eq:probcomputation_rand} of the corresponding SGP encoding is not affine in
parameter $p$ and the probability variable $p_s$ for some state $s\in S$.
More generally, the equality constraints
\eqref{eq:well-defined_probs_rand},
\eqref{eq:probcomputation_rand}, and
\eqref{eq:rewcomputation}
involving $\probmdp$ are not necessarily affine, and thus the SGP may not be a convex program~\cite{boyd_convex_optimization}.
Whereas for convex programs \emph{global optimal solutions} can be
found efficiently~\cite{boyd_convex_optimization}, such guarantees are
not given for SGPs. 
However, we can efficiently obtain local optimal solutions for SGPs in our setting, as shown in the following sections.

\section{Convexification}\label{sec:geometric}
We investigate how to transform the SGP~\eqref{eq:min_rand}--\eqref{eq:rewcomputation} into a convex program by relaxing equality constraints and a lifting of variables of the SGP. 
A certain subclass of SGPs called \emph{geometric programs} (GPs) can be transformed into convex programs~\cite[\S{}2.5]{boyd2007tutorial} and solved efficiently.
A GP is an SGP of the form~\eqref{eq:nl_obj}--\eqref{eq:nl_eq} where $f,g_i\in \posys[\Var]$ and $h_j\in\monos[\Var]$.
We will refer to a constraint with posynomial or monomial function as a posynomial or monomial constraint, respectively.
%
%
%
\subsection{Transformation and relaxation of equality constraints}
As discussed before, the SGP~\eqref{eq:min_rand}--\eqref{eq:rewcomputation} is not convex because of the presence of non-affine equality constraints. 
First observe the following transformation~\cite{boyd2007tutorial}:
\begin{align}
	f\leq h\Longleftrightarrow \frac{f}{h}\leq 1,\label{eq:geo_trans}
\end{align}
for $f\in\posys[\Var]$ and $h\in\monos [\Var]$. Note that monomials are strictly positive (Def.~\ref{def:posy}). This \emph{(division-)transformation} of $f\leq h$ yields a \emph{posynomial inequality constraint}.

We \emph{relax} all equality constraints of SGP~\eqref{eq:min_rand}--\eqref{eq:rewcomputation} that are not monomials to inequalities, then we apply the division-transformation wherever possible. Constraints \eqref{eq:lambda}, \eqref{eq:kappa}, \eqref{eq:well-defined_sched_rand}, \eqref{eq:well-defined_probs_rand}, \eqref{eq:probcomputation_rand}, and \eqref{eq:rewcomputation} are transformed to
\begin{align}
							 &\quad \frac{p_{\sinit}}{\lambda}\leq 1\label{eq:lambda_geo},\\
							 &\quad \frac{c_{\sinit}}{\kappa}\leq 1\label{eq:kappa_geo},\\
			\forall s\in S.	&\quad \sum_{\act\in\Act(s)}\sched^{s,\act}\leq 1\label{eq:welldefined_sched_rand_geo},\\			
			\forall s\in S\, \forall\act\in\Act(s).	 &\quad \sum_{s'\in S}\probmdp(s,\act,s')\leq 1\label{eq:well-defined_probs_rand_geo},\\
			\forall s\in S\setminus T.	&\quad \frac{\sum\limits_{\act\in\Act(s)}\sigma^{s,\act}\cdot\sum\limits_{s'\in S}	\probmdp(s,\act,s')\cdot p_{s'}}{p_s}\leq 1\label{eq:probcomputation_rand_geo},\\
			\forall s\in S\setminus G.	&\quad \frac{\sum\limits_{\act\in\Act(s)} \sigma^{s,\act} \cdot \Bigl(c(s,\act) + \sum\limits_{s'\in S}	\probmdp (s,\act,s') \cdot c_{s'}\Bigr)}{c_s}\leq 1 \label{eq:rewcomputation_geo}.
		\end{align}
These constraints are not necessarily posynomial inequality constraints because (as in Def.~\ref{def:pmdp}) we allow signomial expressions in the transition probability function $\pmdp$. Therefore, replacing \eqref{eq:lambda}, \eqref{eq:kappa}, \eqref{eq:well-defined_sched_rand}, \eqref{eq:well-defined_probs_rand}, \eqref{eq:probcomputation_rand}, and \eqref{eq:rewcomputation} in the SGP with~\eqref{eq:lambda_geo}--\eqref{eq:rewcomputation_geo} does not by itself convert the SGP to a GP.

\subsection{Convexification by lifting}\label{sec:lifting}
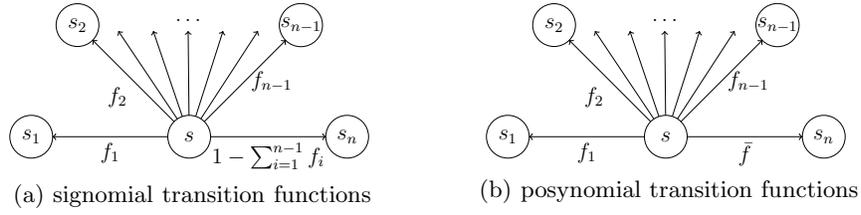
\begin{figure}[t]\centering
	\subfigure[signomial transition functions]{
		\scalebox{0.7}{\begin{tikzpicture}[scale=1, die/.style={inner sep=0,outer sep=0},nodestyle/.style={draw,circle},baseline=(s0)]
    \node [die,state] (s0) at (0,0) {\large$s$};    
    \node [die] (s1) [state, on grid, left=3cm of s0] {\large$s_1$};
    \node [die] (s2) [state, on grid, above left=3cm of s0] {\large$s_2$};
    \node [die] (s3) [state, on grid, above right=3cm of s0] {\large$s_{n-1}$};
    \node [die] (s4) [state, on grid, right=3cm of s0] {\large$s_n$};
    \draw (s0) edge[->] node[auto] {\large$f_1$} (s1);
    \draw (s0) edge[->] node[auto] {\large$f_2$} (s2);
    \draw (s0) edge[->] node[right] {\large$f_{n-1}$} (s3);
    \draw (s0) edge[->] node[below] {\large$1-\sum_{i=1}^{n-1} f_i$} (s4);
	\node [] (dummy1) [on grid, right=0.7cm of s2] {};
	\node [] (dummy2) [on grid, right=0.7cm of dummy1] {};
	\node [] (dummy3) [on grid, right=0.7cm of dummy2] {};
	\node [] (dummy4) [on grid, right=0.7cm of dummy3] {};
	\node [] (dummy5) [on grid, right=0.7cm of dummy4] {};
    \node [] (transitiondots) [on grid, above=.1cm of dummy3] {\large$\ldots$};
    \draw (s0) edge[->] (dummy1);
	\draw (s0) edge[->] (dummy2);	
	\draw (s0) edge[->] (dummy3);
	\draw (s0) edge[->] (dummy4);
	\draw (s0) edge[->] (dummy5);
    \node[draw=white, rectangle, fit=(current bounding box), inner sep=3pt, outer sep=0pt] {};    
\end{tikzpicture}}
	\label{fig:pmc_signomials}
	}\hspace{1cm}
	\subfigure[posynomial transition functions]{
		\scalebox{0.7}{\begin{tikzpicture}[scale=1, die/.style={inner sep=0,outer sep=0},nodestyle/.style={draw,circle},baseline=(s0)]
    \node [die,state] (s0) at (0,0) {\large$s$};    
    \node [die] (s1) [state, on grid, left=3cm of s0] {\large$s_1$};
    \node [die] (s2) [state, on grid, above left=3cm of s0] {\large$s_2$};
    \node [die] (s3) [state, on grid, above right=3cm of s0] {\large$s_{n-1}$};
    \node [die] (s4) [state, on grid, right=3cm of s0] {\large$s_n$};
    \draw (s0) edge[->] node[auto] {\large$f_1$} (s1);
    \draw (s0) edge[->] node[auto] {\large$f_2$} (s2);
    \draw (s0) edge[->] node[right] {\large$f_{n-1}$} (s3);
    \draw (s0) edge[->] node[below] {\large$\bar f$} (s4);
	\node [] (dummy1) [on grid, right=0.7cm of s2] {};
	\node [] (dummy2) [on grid, right=0.7cm of dummy1] {};
	\node [] (dummy3) [on grid, right=0.7cm of dummy2] {};
	\node [] (dummy4) [on grid, right=0.7cm of dummy3] {};
	\node [] (dummy5) [on grid, right=0.7cm of dummy4] {};
    \node [] (transitiondots) [on grid, above=.1cm of dummy3] {\large$\ldots$};
    \draw (s0) edge[->] (dummy1);
	\draw (s0) edge[->] (dummy2);	
	\draw (s0) edge[->] (dummy3);
	\draw (s0) edge[->] (dummy4);
	\draw (s0) edge[->] (dummy5);
    \node[draw=white, rectangle, fit=(current bounding box), inner sep=3pt, outer sep=0pt] {};    
\end{tikzpicture}}
	\label{fig:pmc_monomials}
	}
\caption{Lifting of signomial transition probability function.}
\end{figure}

The relaxed equality constraints~\eqref{eq:well-defined_probs_rand_geo}--\eqref{eq:rewcomputation_geo} involving $\probmdp$ are signomial, rather than posynomial, because the parameters enter Problem~\ref{prob:pmdpsyn} in signomial form. Specifically, consider the relaxed equality constraint~\eqref{eq:probcomputation_rand_geo} at $s_0$ in Example~\ref{ex:die},
\begin{align}
	\label{eq:kys0_relaxed}
	\frac{p\cdot p_{s_1} + (1-p)\cdot p_{s_2}}{p_{s_0}} \leq 1.
\end{align}
The term $(1-p)\cdot p_{s_2}$ is signomial in $p$ and $p_{s_2}$. We \emph{lift} by introducing a new variable $\bar p = 1-p$, and rewrite~\eqref{eq:kys0_relaxed} as a posynomial inequality constraint and an equality constraint in the lifted variables:
\begin{align}
	\frac{p\cdot p_{s_1} + \bar{p} \cdot p_{s_2}}{p_{s_0}} \leq 1,
	\quad
	\bar p = 1-p.
\end{align}
We relax the (non-monomial) equality constraint to $p + \bar p \leq 1$. 
More generally, we restrict the way parameters occur in $\probmdp$ as follows. Refer to Fig.~\ref{fig:pmc_signomials}. For every state $s\in S$ and every action $\act\in\Act(s)$ we require that there exists at most one state $\bar s\in S$ such that $\probmdp(s,\act,\bar s)\in\signomis[\Paramvar]$ and $\probmdp(s,\act,s')\in\posys[\Paramvar]$ for all $s'\in S\setminus\{\bar s\}$. In particular, we require that
	\begin{align*}
		\underbrace{\probmdp(s,\act,\bar s)}_{\in\signomis[\Paramvar]} = 1 - \sum_{s'\in S\setminus\{\bar s\}} \underbrace{\probmdp({s,\act,s'})}_{\in\posys[\Paramvar]}\ .
	\end{align*}
This requirement is met by all benchmarks available at the \tool{PARAM}--webpage~\cite{param_website}.
In general, we lift by introducing a new variable $\bar p_{s,\act,\bar s}=\probmdp(s,\act,\bar s)$ for each such state $s\in S$; refer to Fig.~\ref{fig:pmc_monomials}.
We denote this set of \emph{lifting variables} by $L$. Lifting as explained above then creates a new transition probability function $\bar \probmdp$ where for every $s,s'\in S$ and $\act\in \Act$ we have $\bar\probmdp(s,\act,s')\in\posys[\Paramvar\cup L]$. 

We call the set of constraints obtained through transformation, relaxation, and lifting of every constraint of the SGP~\eqref{eq:lambda}--\eqref{eq:rewcomputation} as shown above the \emph{convexified constraints}. 
Any posynomial objective subject to the convexified constraints forms by construction a GP over the pMDP parameters $\Paramvar$, the SGP additional variables $W$, and the lifting variables $L$. 

\subsection{Tightening the constraints}
A solution of the GP as obtained in the previous section does not have a direct relation to the original SGP~\eqref{eq:min_rand}--\eqref{eq:rewcomputation}.
In particular, a solution to the GP may not have the relaxed constraints satisfied with equality. 
For \eqref{eq:welldefined_sched_rand_geo} and~\eqref{eq:well-defined_probs_rand_geo}, the induced parameter valuation and the scheduler are not well-defined, \ie, the probabilities may not sum to one. \sj{In particular, this makes it trivial to fulfill the GP, just assign 0 to all schedulers}
We need to relate the relaxed and lifted GP to Problem~\ref{prob:pmdpsyn}. By defining a \emph{regularization function} $F$ over all parameter and scheduler variables, we ensure that the constraints are satisfied with equality; enforcing well-defined probability distributions.
%
\begin{align}
	F = \sum_{p\in \Paramvar} \frac{1}{p} + \sum_{\bar p\in L} \frac{1}{\bar p} + \sum_{s\in S,\act\in\Act(s)} \frac{1}{\sigma_{s,\act}}\label{eq:regularization} \ .
\end{align}
The function $F$ is monotone in all its variables. We discard the original objective $f$ in~\eqref{eq:min_rand} and form a GP with the regularization objective $F$~\eqref{eq:regularization}:
\begin{align}
			\text{minimize } &\quad F\label{eq:min_rand_geofull}\\
			\text{subject to}\notag \\
							 &\quad \frac{p_{\sinit}}{\lambda}\leq 1\label{eq:lambda_geofull},\\
							 &\quad \frac{c_{\sinit}}{\kappa}\leq 1\label{eq:kappa_geofull},\\
			\forall s\in S.	&\quad \sum_{\act\in\Act(s)}\sched^{s,\act}\leq 1\label{eq:welldefined_sched_rand_geofull},\\
			\forall s\in S\, \forall\act\in\Act(s). &\quad \sched^{s,\act} \leq 1			\label{eq:sched_is_dist_geofull},\\					
			\forall s\in S\, \forall\act\in\Act(s).	 &\quad \sum_{s'\in S}\bar\probmdp(s,\act,s')\leq 1\label{eq:well-defined_probs_rand_geofull},\\
				\forall s, s'\in S\, \forall\act\in\Act(s).	 &\quad\bar{\probmdp}(s,\act,s') \leq 1,	\label{eq:probs_is_prob_geofull}\\
	\forall s\in T.	&\quad p_s=1,															\label{eq:targetprob_rand_geofull}\\
			\forall s\in S\setminus T.	&\quad \frac{\sum\limits_{\act\in\Act(s)}\sigma^{s,\act}\cdot\sum\limits_{s'\in S}	\bar\probmdp(s,\act,s')\cdot p_{s'}}{p_s}\leq 1\label{eq:probcomputation_rand_geofull},\\
			\forall s\in S\setminus G.	&\quad \frac{\sum\limits_{\act\in\Act(s)} \sigma^{s,\act} \cdot \Bigl(c(s,\act) + \sum\limits_{s'\in S}	\bar\probmdp (s,\act,s') \cdot c_{s'}\Bigr)}{c_s}\leq 1 \label{eq:rewcomputation_geofull}.
\end{align}
%
Since the objective $F$~\eqref{eq:regularization} and the inequality constraints~\eqref{eq:welldefined_sched_rand_geofull} and~\eqref{eq:well-defined_probs_rand_geofull} are monotone in $\Paramvar$, $L$, and the scheduler variables, each optimal solution for a feasible problem satisfies them with equality. We obtain a well-defined scheduler $\sched$ and a valuation $u$ as in Problem 1. Note that variables from~\eqref{eq:targetrew} are explicitly excluded from the GP by treating them as constants.

The reachability probability constraints~\eqref{eq:probcomputation_rand_geofull} and cost constraints \eqref{eq:rewcomputation_geofull} need not be satisfied with equality. However, \eqref{eq:probcomputation_rand_geofull} is equivalent to 
\begin{align*}
	{p_s} \geq {\sum\limits_{\act\in\Act(s)}\sigma^{s,\act}\cdot\sum\limits_{s'\in S} \bar\probmdp(s,\act,s')\cdot p_{s'}}
\end{align*}
for all $s\in S\setminus T$ and $\act\in \Act$.
The probability variables $p_s$ are assigned upper bounds on the actual probability to reach the target states $T$ under scheduler $\sched$ and valuation $u$. Put differently, the $p_s$ variables cannot be assigned values that are lower than the actual probability; ensuring that $\sched$ and $u$ induce satisfaction of the specification given by~\eqref{eq:lambda_geofull} if the problem is feasible and $\sched$ and $u$ are well-defined. An analogous reasoning applies to the expected cost computation~\eqref{eq:rewcomputation_geofull}. 
A solution consisting of a scheduler or valuation that are not well-defined occurs only if Problem~\ref{prob:pmdpsyn} itself is infeasible. Identifying that such a solution has been obtained is easy.
These facts allow us to state the main result of this section.
\begin{theorem}
	\label{thm:main}
	A solution of the GP~\eqref{eq:min_rand_geofull}--\eqref{eq:rewcomputation_geofull} inducing well-defined scheduler $\sched$ and valuation $u$ is a feasible solution to Problem 1.
\end{theorem}
Note that the actual probabilities induced by $\sched$ and $u$ for the given pMDP $\mdp$ are given by the MC $\mdp^\sched[u]$ induced by $\sched$ and instantiated by $u$.
Since all variables are implicitly positive in a GP, no transition probability function will be instantiated to probability zero. 
The case of a scheduler variable being zero to induce the optimum can be excluded by a previous graph analysis.

\section{Sequential Geometric Programming}\label{sec:approximation}

We showed how to efficiently obtain a feasible solution
 for Problem~\ref{prob:pmdpsyn} by solving GP~\eqref{eq:min_rand_geofull}--\eqref{eq:rewcomputation_geofull}. 
%
We propose a \emph{sequential convex programming} trust-region method to compute a local optimum of the SGP~\eqref{eq:min_rand}--\eqref{eq:rewcomputation}, following~\cite[\S{}9.1]{boyd2007tutorial}, solving a sequence of GPs. We obtain each GP by replacing signomial functions in equality constraints of the SGP~\eqref{eq:min_rand}--\eqref{eq:rewcomputation} with \emph{monomial approximations} of the functions.

\begin{definition}[Monomial approximation]
Given a posynomial $f \in \signomis[\Var]$, variables $\Var = \{x_1,\dots,x_n\}$, and a valuation $u \in \valuations^\Var$, a \emph{monomial approximation $\hat f\in\monos[\Var]$ for $f$ near $u$} is 
\begin{align*}
	\forall i. 1 \leq i \leq n
	\quad
	\hat f = f[u]
	\prod_{i=1}^{n}\Bigg(\dfrac{x_{i}}{u(x_i)}\Bigg)^{a_{i}},
	\quad
	\text{where }
	a_{i}=\dfrac{u(x_i)}{f[u]}\dfrac{\partial f}{\partial x_{i}}[u].
\end{align*}	
\end{definition}
Intuitively, we compute a \emph{linearization} $\hat f$ of $f\in\signomis[\Var]$ around a fixed valuation $u$.
%
%
We enforce the fidelity of monomial approximation $\hat{f}$ of $f \in \signomis[\Var]$ by restricting valuations to remain within a set known as \emph{trust region}. We define the following constraints on the variables $\Var$ with $t > 1$ determining the size of the trust region:
\begin{align}
	\label{eq:trust_region}
	\forall i. 1 \leq i \leq n
	\quad (\nicefrac{1}{t})\cdot u(x_i) \leq x_i \leq t\cdot u(x_i)
\end{align}
%
%
%
%
%
%
For a given valuation $u$, we approximate the SGP~\eqref{eq:min_rand}--\eqref{eq:rewcomputation} to obtain a \emph{local GP} as follows. 
First, we apply a \emph{lifting} procedure (Section~\ref{sec:lifting}) to the SGP ensuring that all constraints consist of posynomial functions. The thus obtained posynomial inequality constraints are included in the local GP.
After replacing posynomials in every equality constraint by their monomial approximations near $u$, the resulting monomial equality constraints are also included.
Finally, we add trust region constraints~\eqref{eq:trust_region} for scheduler and parameter variables. The objective function is the same as for the SGP.
The optimal solution of the local GP is not necessarily a feasible solution to the SGP. 
Therefore, we first normalize the scheduler and parameter values to obtain well-defined probability distributions. These normalized values are used to compute precise probabilities and expected cost using \tool{PRISM}. The steps above provide a feasible solution of the SGP.

We use such approximations to obtain a sequence of feasible solutions to the SGP approaching a local optimum of the SGP. 
First, we compute a feasible solution $u^{(0)}$ for Problem~\ref{prob:pmdpsyn} (Section \ref{sec:geometric}), 
%
forming the initial point of a sequence of solutions $u^{(0)},\ldots, u^{(N)}, N \in \N$. 
The solution $u^{(k)}$ for $0\leq k\leq N$ is obtained from a local GP defined using $u^{(k-1)}$ as explained above.



The parameter $t$ for each iteration $k$ is determined based on its value for the previous iteration, and the ratio of $ f\left[u^{(k-1)}\right] $ to $f\left[u^{(k-2)}\right]$, where $f$ is the objective function in \eqref{eq:min_rand}. The iterations are stopped when $\left| f\left[u^{(k)}\right] - f\left[u^{(k-1)}\right] \right| <  \epsilon$. Intuitively, $\epsilon$ defines the required improvement on the objective value for each iteration; once there is not enough improvement the process terminates.

\section{Applications}\label{sec:applications}
We discuss two applications and their restrictions for the general SGP~\eqref{eq:min_rand}--\eqref{eq:rewcomputation}.\vspace{-.2cm}
\paragraph{Model repair.}
For MC $\dtmc$ and specification $\varphi$ with $\dtmc\not\models\varphi$, the \emph{model repair} problem~\cite{bartocci2011model} is to transform $\dtmc$ to $\dtmc'$ such that $\dtmc'\models\varphi$. The transformation involves a change of transition probabilities. Additionally, a cost function measures the change of probabilities. The natural underlying model is a pMC where parameters are added to probabilities. The cost function is minimized subject to constraints that induce satisfaction of $\varphi$. In~\cite{bartocci2011model}, the problem is given as NLP. Heuristic~\cite{pathak-et-al-nfm-2015} and simulation-based methods~\cite{chen2013model} (for MDPs) were presented.

Leveraging our results, one can readily encode model repair problems for MDPs, multiple objectives, and restrictions on probability or cost changes directly as NLPs. The encoding as in~\cite{bartocci2011model} is  handled by our method in Section~\ref{sec:approximation} as it involves signomial constraints. We now propose a more efficient approach, which encodes the change of probabilities using monomial functions.
Consider an MDP $\MdpInit$ and specifications $\varphi_1,\ldots,\varphi_q$ with $\mdp\not\models\varphi_1\land\ldots\land\varphi_q$. For each probability $\pmdp(s,\act,s')=a\in\R$ that may be changed, introduce a parameter $p$, forming the parameter set $\Paramvar$. We define a parametric transition probability function by $\pmdp'(s,\act,s')=p\cdot a\in\monos[V]$. The quadratic cost function is for instance  $f=\sum_{p\in V} p^2\in\posys[V]$. 
By minimizing the sum of squares of the parameters (with some regularization), the change of probabilities is minimized.

By incorporating these modifications into SGP~\eqref{eq:min_rand}--\eqref{eq:rewcomputation}, our approach is directly applicable. Either we restrict the cost function $f$ to an upper bound, and efficiently solve a feasibility problem (Section~\ref{sec:geometric}), or we compute a local minimum of the cost function (Section~\ref{sec:approximation}).
In contrast to~\cite{bartocci2011model}, our approach works for MDPs and has an efficient solution. While~\cite{chen2013model} uses fast simulation techniques, we can directly incorporate multiple objectives and restrictions on the results while offering an efficient numerical solution of the problem.

\paragraph{Parameter space partitioning.}
For pMDPs, 
tools like \tool{PRISM}~\cite{KNP11} or \tool{PROPhESY}~\cite{dehnert-et-al-cav-2015} aim at partitioning the parameter space into regions with respect to a specification.
A \emph{parameter region} is given by a convex polytope defined by linear inequalities over the parameters, restricting valuations to a region. Now, for pMDP $\mdp$ a region is \emph{safe} regarding a specification $\varphi$, if no valuation $u$ inside this region and no scheduler $\sched$ induce $\mdp^\sched[u]\not\models\varphi$. Vice versa, a region is unsafe, if there is no valuation and scheduler such that the specification is satisfied. 
In~\cite{dehnert-et-al-cav-2015}, this certification is performed using SMT solving.
More efficiency is achieved by using an approximation method~\cite{quatmann-et-al-atva-2016}.

Certifying regions to be unsafe is directly possible using our approach.
Assume pMDP $\mdp$, specifications $\varphi_1,\ldots,\varphi_q$, and a region candidate defined by a set of linear inequalities.
We incorporate the inequalities in the NLP~\eqref{eq:min_rand}--\eqref{eq:rewcomputation}. If the feasibility problem (Section~\ref{sec:geometric}) has no solution, the region is unsafe. This yields the \emph{first efficient numerical method} for this problem of which we are aware.
Proving that a region is safe is more involved. Given one specification $\varphi=\reachProplT$, we maximize the probability to reach $T$. If this probability is at most $\lambda$, the region is safe. For using our method from Section~\ref{sec:approximation}, one needs domain specific knowledge to show that a local optimum is a global optimum.
%


\section{Experiments}\label{sec:experiments}
We implemented a prototype using the \tool{Python} interfaces of the probabilistic model checker \tool{STORM}~\cite{DBLP:journals/corr/DehnertJK016} and the optimization solver \tool{MOSEK}~\cite{mosek}.
 All experiments were run on a 2.6 GHz machine with 32 GB RAM. 
 We used \tool{PRISM}~\cite{KNP11} to correct approximation errors as explained before. 
We evaluated our approaches using mainly examples from the \tool{PARAM}--webpage~\cite{param_website} and from \tool{PRISM}~\cite{KNP12b}.
We considered several parametric instances of the \emph{Bounded Retransmission Protocol} (BRP)~\cite{HSV94},  \emph{NAND Multiplexing}~\cite{HJ02}, and the \emph{Consensus} protocol (CONS)~\cite{consensus}. For BRP, we have a pMC and a pMDP version, NAND is a pMC, and CONS is a pMDP.
For obtaining feasibility solutions, we compare to the SMT solver \tool{Z3}~\cite{demoura_nlsat}. For additional optimality criteria, there is no comparison to another tool possible as \tool{IPOPT}~\cite{ipopt} already fails for the smallest instances we consider.

\begin{figure}[t]
	\centering
	\subfigure[Benchmark results]{\scalebox{0.75}{\begin{tabular}[b]{@{}crrlrlrr@{}}
\toprule
Benchmark       & \#{}states & \#{}par     & specs      & \tool{MOSEK} (s)  & & Z3 \\
\midrule
BRP (pMC)   & $5382$   &  $2$    & $\er,\p,*$    & $23.17$ &($6.48$) & $-$                   \\
   & $112646$   &  $2$    & $\er,\p,*$    & $3541.59$  &($463.74$) & $-$      \\
      & $112646$   &  $4$    & $\er,\p,*$    & $4173.33$  &($568.79$) & $-$              \\

  & $5382$   &  $2$    & $\er,\p$    & $3.61$   & &$904.11$                   \\
  & $112646$   &  $2$    & $\er,\p$    & $479.08$  & &$\mathit{TO}$                  \\
  
NAND (pMC)  & $4122$   &  $2$    & $\er,\p,*$    & $14.67$ &($2.51$) & $-$                   \\
   & $35122$   &  $2$    & $\er,\p,*$    & $1182.41$ &($95.19$) & $-$                  \\
   & $4122$   &  $2$    & $\er,\p$    & $1.25$   && $1.14$                   \\
  & $35122$   &  $2$    & $\er,\p$    & $106.40$  && $11.49$                  \\
BRP (pMDP)   & $5466$   &  $2$    & $\er,\p,*$    & $31.04$ &($8.11$) & $-$                   \\      
   & $112846$   &  $2$    & $\er,\p,*$    & $4319.16$  &($512.20$) & $-$                   \\    
   & $5466$   &  $2$    & $\er,\p$    & $4.93$  && $1174.20$                   \\      
   & $112846$   &  $2$    & $\er,\p$    & $711.50$   && $\mathit{TO}$                   \\  
CONS (pMDP)  & $4112$   &  $2$    & $\er,\p,*$    & $102.93$  &($1.14$) & $-$                   \\      
   & $65552$   &  $2$    & $\er,\p,*$    & $\mathit{TO}$  &&  $-$                   \\    
   & $4112$   &  $2$    & $\er,\p$    & $6.13$   && $\mathit{TO}$                   \\      
   & $65552$   &  $2$    & $\er,\p$    & $1361.96$  & & $\mathit{TO}$                   \\  
\bottomrule\end{tabular}
}\label{tab:scalability}}
	\subfigure[Sensitivity to \#par]{\scalebox{0.95}{\pgfplotsset{every axis/.append style={
                    legend style={font=\tiny, at={((0.5,-0.19))}, align=left, anchor=north,draw=none ,mark size=2pt},
                    }}%
\pgfplotsset{footnotesize}
 \begin{tikzpicture}
\draw[black, use as bounding box] (-1.2,-1.6) rectangle (2.7,4.2);
 \iftrue
\begin{axis}[width=3.9cm, height=5.5cm, ylabel={Time (s)}, xlabel={Number of parameters},
axis x line=bottom,mark size=0.8pt,
 axis y line=left,
 ymode = log,
 ymin = 0,
 ymax = 100,
 ytick = {0.1,0.2,0.5,1,2,5,10,20,50,100},
 yticklabels = {0.1,0.2,0.5,1,2,5,10,20,50,$\mathit{TO}$},
 xtick={2,3,4,5,6,7,8},
 x label style={font=\scriptsize,at={(axis description cs:0.5,0.05)},anchor=north},
 y label style={font=\scriptsize, at={(axis description cs:0.25,.5)},anchor=south},
 legend columns=2,
 legend entries={MOSEK,Z3,PROPhESY},
 legend image post style={scale=0.5},
legend cell align=left,
 yticklabel style={font=\tiny},
 xticklabel style={font=\tiny}]
]
    \addplot+[mark=*] file {data/die_convex.dat};
    \addplot+[mark=square] file {data/die_smt.dat};
    \addplot+[mark=square*] file {data/die_stateelim.dat};
\end{axis}  
\fi
\end{tikzpicture}\label{plot:param}}}
\caption{Experiments.}
\end{figure}
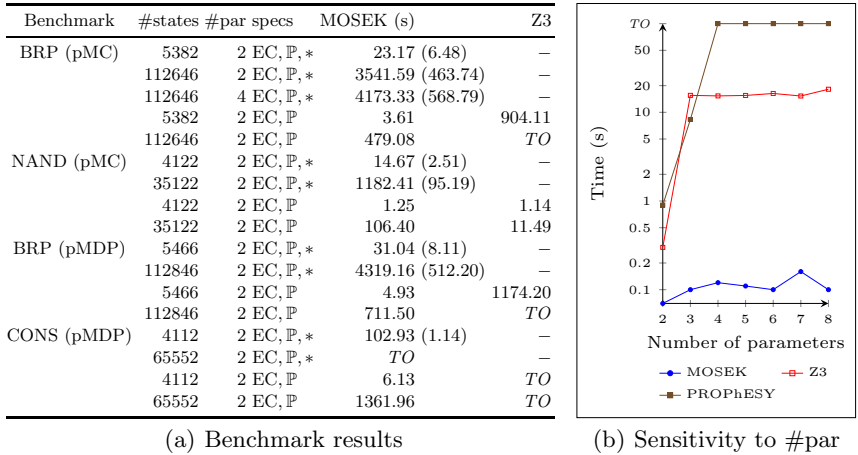

Fig.~\ref{tab:scalability} states for each benchmark instance the number of states (\#{}states) and the number of parameters (\#{}par). We defined two specifications consisting of a expected cost property ($\er$) and a reachability property ($\p$). For some benchmarks, we also maximized the probability to reach a set of ``good states'' ($*$). We list the times taken by \tool{MOSEK}; for optimality problems we also list the times \tool{PRISM} took to compute precise probabilities or costs (Section~\ref{sec:approximation}). For feasibility problems we list the times of \tool{Z3}. The timeout (\emph{TO}) is $90$ minutes.
%
%

We observe that both for feasibility with optimality criteria we can handle most benchmarks of up to $10^5$ states within the timeout, while we ran into a timeout for CONS. The number of iterations $N$ in the sequential convex programming is less than $12$ for all benchmarks with $\epsilon=10^{-3}$.
As expected, simply solving feasibility problems is faster by at least one order of magnitude. Raising the number of parameters from $2$ to $4$ for BRP does not cause a major performance hit, contrary to existing tools. For all benchmarks except NAND, \tool{Z3} only delivered results for the smallest instances within the timeout. 

To demonstrate the insensitivity of our approach to the number of parameters, we considered a pMC of rolling multiple  Knuth--Yao dice with $156$ states, $522$ transitions and considered instances with up to $8$ different parameters. The timeout is $100$ seconds.
In Fig.~\ref{plot:param} we compare our encoding in \tool{MOSEK} for this benchmark to the mere computation of a rational function using \tool{PROPhESY}~\cite{dehnert-et-al-cav-2015} and again to \tool{Z3}. \tool{PROPhESY} already runs into a timeout for $4$ parameters\footnote{Due to the costly computation of greatest common divisors employed in \prophesy.}.
\tool{Z3} needs around $15$ seconds for most of the tests. Using GPs with \tool{MOSEK} proves far more efficient  as it needs less than one second for all instances.

In addition, we test model repair (Section~\ref{sec:applications}) on a BRP instance with $17415$ states for $\varphi=\reachProp{0.9}{T}$. The initial parameter instantiation violates $\varphi$. We performed model repair towards satisfaction of $\varphi$. The probability of reaching $T$ results in $0.79$ and the associated cost is $0.013$. The computation time is $21.93$ seconds. We compare our result to an implementation of~\cite{chen2013model}, where the probability of reaching $T$ is $0.58$ and the associated cost is $0.064$. However, the time for the simulation-based method is only $2.4$ seconds, highlighting the expected trade-off between optimality and computation times for the two methods.
	
Finally, we encode model repair for the small pMC from Example~\ref{ex:die} in \tool{IPOPT}, see~\cite{bartocci2011model}. For $\psi=\reachProp{0.125}{T}$ where $T$ represents the outcome of the die being $2$, the initial instantiation induces probability $1/6$. With our method, the probability of satisfying $\psi$ is $0.1248$ and the cost is $0.0128$. With \tool{IPOPT}, the probability is $0.125$ with cost $0.1025$, showing that our result is nearly optimal.

\section{Conclusion and future work}\label{sec:conclusion}
We presented a way to use convex optimization in the field of parameter synthesis for parametric Markov models. Using our results, many NLP encodings of related problems now have a direct and efficient solution.

Future work will concern the integration of these methods into mature tools like \tool{PRISM} or \tool{PROPhESY} to enable large-scale benchmarking by state space reduction techniques and advanced data structures. Moreover, we will explore extensions to richer models like continuous-time Markov chains~\cite{katoen2016probabilistic}.

\newpage
\bibliographystyle{plainyr}
\bibliography{literature}

\begin{thebibliography}{10}

\bibitem{DBLP:journals/ior/SatiaL73}
Jay~K. Satia and Roy E.~Lave Jr.
\newblock Markovian decision processes with uncertain transition probabilities.
\newblock {\em Operations Research}, 21(3):728--740, 1973.

\bibitem{KY76}
Donald~E. Knuth and Andrew~C. Yao.
\newblock The complexity of nonuniform random number generation.
\newblock In Joseph~F. Traub, editor, {\em Algorithms and Complexity: New
  Directions and Recent Results}, page 375. Academic Press, 1976.

\bibitem{consensus}
James Aspnes and Maurice Herlihy.
\newblock Fast randomized consensus using shared memory.
\newblock {\em Journal of Algorithms}, 15(1):441--460, 1990.

\bibitem{HSV94}
L.~Helmink, M.~Sellink, and F.~Vaandrager.
\newblock Proof-checking a data link protocol.
\newblock In {\em TYPES}, volume 806 of {\em LNCS}, pages 127--165. Springer,
  1994.

\bibitem{Las01}
Jean~B. Lasserre.
\newblock Global optimization with polynomials and the problem of moments.
\newblock {\em {SIAM} Journal on Optimization}, 11(3):796--817, 2001.

\bibitem{HJ02}
Jie Han and Pieter Jonker.
\newblock A system architecture solution for unreliable nanoelectronic devices.
\newblock {\em IEEE Transactions on Nanotechnology}, 1:201--208, 2002.

\bibitem{boyd2007tutorial}
Stephen Boyd, Seung-Jean Kim, Lieven Vandenberghe, and Arash Hassibi.
\newblock A tutorial on geometric programming.
\newblock {\em Optimization and Engineering}, 8(1), 2007.

\bibitem{boyd2008sequential}
Stephen Boyd.
\newblock Sequential convex programming.
\newblock Lecture Notes, 2008.

\bibitem{DBLP:journals/lmcs/EtessamiKVY08}
Kousha Etessami, Marta Kwiatkowska, Moshe~Y. Vardi, and Mihalis Yannakakis.
\newblock Multi-objective model checking of {M}arkov decision processes.
\newblock {\em LMCS}, 4(4), 2008.

\bibitem{ipopt}
Lorenz~T. Biegler and Victor~M. Zavala.
\newblock Large-scale nonlinear programming using {IPOPT}: {A}n integrating
  framework for enterprise-wide dynamic optimization.
\newblock {\em Computers {\&} Chemical Engineering}, 33(3):575--582, 2009.

\bibitem{PARAM10}
Ernst~Moritz Hahn, Holger Hermanns, Bj{\"o}rn Wachter, and Lijun Zhang.
\newblock {PARAM}: A model checker for parametric {M}arkov models.
\newblock In {\em CAV}, volume 6174 of {\em LNCS}, pages 660--664. Springer,
  2010.

\bibitem{param_sttt}
Ernst~Moritz Hahn, Holger Hermanns, and Lijun Zhang.
\newblock Probabilistic reachability for parametric {M}arkov models.
\newblock {\em STTT}, 13(1):3--19, 2010.

\bibitem{bartocci2011model}
Ezio Bartocci, Radu Grosu, Panagiotis Katsaros, CR~Ramakrishnan, and Scott~A
  Smolka.
\newblock Model repair for probabilistic systems.
\newblock In {\em TACAS}, volume 6605 of {\em LNCS}, pages 326--340. Springer,
  2011.

\bibitem{DBLP:conf/tacas/ForejtKNPQ11}
Vojtech Forejt, Marta Kwiatkowska, Gethin Norman, David Parker, and Hongyang
  Qu.
\newblock Quantitative multi-objective verification for probabilistic systems.
\newblock In {\em TACAS}, volume 6605 of {\em LNCS}, pages 112--127. Springer,
  2011.

\bibitem{KNP11}
Marta Kwiatkowska, Gethin Norman, and David Parker.
\newblock \tool{PRISM} 4.0: Verification of probabilistic real-time systems.
\newblock In {\em CAV}, volume 6806 of {\em LNCS}, pages 585--591. Springer,
  2011.

\bibitem{DBLP:conf/atva/ForejtKP12}
Vojtech Forejt, Marta Kwiatkowska, and David Parker.
\newblock Pareto curves for probabilistic model checking.
\newblock In {\em ATVA}, volume 7561 of {\em LNCS}, pages 317--332. Springer,
  2012.

\bibitem{demoura_nlsat}
Dejan Jovanovic and Leonardo~Mendon\c{c}a de~Moura.
\newblock Solving non-linear arithmetic.
\newblock In {\em IJCAR}, volume 7364 of {\em LNCS}, pages 339--354. Springer,
  2012.

\bibitem{KNP12b}
Marta Kwiatkowska, Gethin Norman, and David Parker.
\newblock The \tool{PRISM} benchmark suite.
\newblock In {\em QEST}, pages 203--204. IEEE CS, 2012.

\bibitem{chen2013model}
Taolue Chen, Ernst~Moritz Hahn, Tingting Han, Marta Kwiatkowska, Hongyang Qu,
  and Lijun Zhang.
\newblock Model repair for {M}arkov decision processes.
\newblock In {\em TASE}, pages 85--92. IEEE CS, 2013.

\bibitem{seshia_et_al_cav_13}
Alberto Puggelli, Wenchao Li, Alberto~L. Sangiovanni-Vincentelli, and Sanjit~A.
  Seshia.
\newblock Polynomial-time verification of {PCTL} properties of {MDPs} with
  convex uncertainties.
\newblock In {\em CAV}, volume 8044 of {\em LNCS}, pages 527--542. Springer,
  2013.

\bibitem{DBLP:conf/cmsb/CeskaDKP14}
Milan Ceska, Frits Dannenberg, Marta Kwiatkowska, and Nicola Paoletti.
\newblock Precise parameter synthesis for stochastic biochemical systems.
\newblock In {\em CMSB}, volume 8859 of {\em LNCS}, pages 86--98. LNCS, 2014.

\bibitem{iscasmc}
Ernst~Moritz Hahn, Yong Li, Sven Schewe, Andrea Turrini, and Lijun Zhang.
\newblock {iscasMc}: {A} web-based probabilistic model checker.
\newblock In {\em FM}, volume 8442 of {\em LNCS}, pages 312--317. Springer,
  2014.

\bibitem{jansen-et-al-qest-2014}
Nils Jansen, Florian Corzilius, Matthias Volk, Ralf Wimmer, Erika \'Abrah\'am,
  Joost-Pieter Katoen, and Bernd Becker.
\newblock Accelerating parametric probabilistic verification.
\newblock In {\em QEST}, volume 8657 of {\em LNCS}, pages 404--420. Springer,
  2014.

\bibitem{rosenblum-et-al-atva-2014}
Guoxin Su and David~S. Rosenblum.
\newblock Nested reachability approximation for discrete-time {M}arkov chains
  with univariate parameters.
\newblock In {\em ATVA}, volume 8837 of {\em LNCS}, pages 364--379. Springer,
  2014.

\bibitem{dehnert-et-al-cav-2015}
Christian Dehnert, Sebastian Junges, Nils Jansen, Florian Corzilius, Matthias
  Volk, Harold Bruintjes, Joost{-}Pieter Katoen, and Erika {\'{A}}brah{\'{a}}m.
\newblock Prophesy: {A} probabilistic parameter synthesis tool.
\newblock In {\em {CAV} {(1)}}, volume 9206 of {\em LNCS}, pages 214--231.
  Springer, 2015.

\bibitem{param_website}
{\tool{PARAM} Website}, 2015.
\newblock \url{http://depend.cs.uni-sb.de/tools/param/}.

\bibitem{pathak-et-al-nfm-2015}
Shashank Pathak, Erika {\'{A}}brah{\'{a}}m, Nils Jansen, Armando Tacchella, and
  Joost{-}Pieter Katoen.
\newblock A greedy approach for the efficient repair of stochastic models.
\newblock In {\em {NFM}}, volume 9058 of {\em LNCS}, pages 295--309. Springer,
  2015.

\bibitem{DBLP:journals/iandc/BortolussiMS16}
Luca Bortolussi, Dimitrios Milios, and Guido Sanguinetti.
\newblock Smoothed model checking for uncertain continuous-time {M}arkov
  chains.
\newblock {\em Inf. Comput.}, 247:235--253, 2016.

\bibitem{DBLP:journals/corr/DehnertJK016}
Christian Dehnert, Sebastian Junges, Joost{-}Pieter Katoen, and Matthias Volk.
\newblock The probabilistic model checker storm (extended abstract).
\newblock {\em CoRR}, abs/1610.08713, 2016.

\bibitem{DBLP:journals/ai/DelgadoBDS16}
Karina~Valdivia Delgado, Leliane~N. de~Barros, Daniel~B. Dias, and Scott
  Sanner.
\newblock Real-time dynamic programming for {M}arkov decision processes with
  imprecise probabilities.
\newblock {\em Artif. Intell.}, 230:192--223, 2016.

\bibitem{katoen2016probabilistic}
Joost-Pieter Katoen.
\newblock The probabilistic model checking landscape.
\newblock In {\em IEEE Symposium on Logic In Computer Science (LICS)}. ACM,
  2016.

\bibitem{DBLP:conf/popl/LongR16}
Fan Long and Martin Rinard.
\newblock Automatic patch generation by learning correct code.
\newblock In {\em POPL}, pages 298--312. {ACM}, 2016.

\bibitem{quatmann-et-al-atva-2016}
Tim Quatmann, Christian Dehnert, Nils Jansen, Sebastian Junges, and
  Joost{-}Pieter Katoen.
\newblock Parameter synthesis for {M}arkov models: Faster than ever.
\newblock In {\em ATVA}, volume 9938 of {\em LNCS}, pages 50--67, 2016.

\bibitem{su-et-al-icse-2016-qosevaluation}
Guoxin Su, David~S. Rosenblum, and Giordano Tamburrelli.
\newblock Reliability of run-time qos evaluation using parametric model
  checking.
\newblock In {\em ICSE}. ACM, 2016.
\newblock to appear.

\bibitem{mosek}
MOSEK ApS.
\newblock {\em The MOSEK optimization toolbox for PYTHON. Version 7.1 (Revision
  60)}, 2015.

\bibitem{BK08}
Christel Baier and Joost-Pieter Katoen.
\newblock {\em Principles of Model Checking}.
\newblock MIT Press, 2008.

\bibitem{bertsekas1999nonlinear}
Dimitri~P. Bertsekas.
\newblock {\em Nonlinear Programming}.
\newblock Athena Scientific Belmont, 1999.

\bibitem{boyd_convex_optimization}
Stephen Boyd and Lieven Vandenberghe.
\newblock {\em Convex Optimization}.
\newblock Cambridge University Press, New York, NY, USA, 2004.

\bibitem{thrun2005probabilistic}
Sebastian Thrun, Wolfram Burgard, and Dieter Fox.
\newblock {\em Probabilistic Robotics}.
\newblock MIT Press, 2005.

\end{thebibliography}

\end{document}